\documentclass[%
 reprint,
 amsmath,amssymb,
 aps,
]{revtex4-1}

\usepackage{graphicx}
\usepackage{dcolumn}
\usepackage{bm}
\usepackage[
      colorlinks=true,
      linkcolor=blue,
      urlcolor=blue,
      filecolor=black,
      citecolor=blue,
            ]{hyperref}

\begin{document}


\title {On the critical behavior of gapped gravitational collapse in confined spacetime}

\author{Rong-Gen Cai$^{1,2}$}
 \email{cairg@itp.ac.cn}
\author{Li-Wei Ji$^1$}
 \email{jiliwei@itp.ac.cn}
 \author{Run-Qiu Yang$^3$}
 \email{aqiu@kias.re.kr}
\affiliation{$^1$CAS Key Laboratory of Theoretical Physics, Institute of Theoretical Physics,
	 Chinese Academy of Sciences, Beijing 100190, China  \\
	$^2$ Center for Gravitational Physics, Yukawa Institute for Theoretical Physics, 
 Kyoto University, Kyoto 606-8502, Japan \\
 $^3$ Quantum Universe Center, Korea Institute for Advanced Study, Seoul 130-722, Korea
	}

\begin{abstract}
The gravitational collapse of a massless scalar field enclosed with a perfectly reflecting wall in a spacetime with  a cosmological constant $\Lambda$ is investigated. The mass scaling for the gapped collapse  $ M_{AH}-M_g \propto (\epsilon_c-\epsilon)^\xi$  is confirmed and a new time scaling for the gapped collapse $T_{AH}-T_g\propto(\epsilon_c-\epsilon)^\zeta$ is found. We find that  both of these two critical exponents depend on the  combination $\Lambda R^2$, where $R$ is the radial position of the  reflecting wall. Especially, we find an evolution of the critical exponent $\xi$ from $0.37$ in the confined asymptotic dS case  with $\Lambda R^2=1.5$ to  $0.7$   in asymptotic AdS case ($\Lambda R^2\rightarrow-\infty$),  while the critical exponent $\zeta$ varies from $0.10$ to $0.26$, which shows the new critical behavior for the gapped  collapse is  essentially different from the one in the Choptuik's case.
\end{abstract}

\maketitle


\section{\label{sec:level1_1}Introduction}
In  1993,  Choptuik~\cite{Choptuik:1992jv} found an interesting so-called type \uppercase\expandafter{\romannumeral2} critical phenomenon  in the gravitational collapse of a massless scalar field in spherically symmetric asymptotically flat spacetimes.  He looked closely at the threshold between black hole(BH) formation and dispersion, and found that the masses of BHs formed from supercritical configurations follow a power-law behavior: $M_{BH}\propto(\epsilon-\epsilon_0)^\gamma$. Here $\epsilon$ parameterizes a one-parameter family of initial data, $\epsilon_0$ corresponds to the threshold and $\gamma$ is a universal exponent which is around $0.37$. A few years later, Choptuik, Chmaj and Bizon found the type \uppercase\expandafter{\romannumeral1} critical phenomenon  in gravitational collapse of Yang-Mills field \cite{Choptuik:1996yg}. The BH formations here turned on at a finite mass (mass gap), unlike the type \uppercase\expandafter{\romannumeral2} case where the BH formation turned on at zero mass. The BH masses of supercritical solutions in this case do not follow a power-law scaling. Instead, the span of time which describes how long the configuration stays in the vicinity of the critical solution scales as: $T\propto \ln (\epsilon-\epsilon_0)$.
 For recent reviews on the critical phenomenon in gravitational collapse,  please see \cite{Gundlach:2002sx,Gundlach:2007gc}.

When it comes to the asymptotically AdS spacetime, the situation is very different. Due to the confinement property of the timelike boundary of AdS, the subcritical configurations can be reflected by the boundary and also collapse into BHs. There are many thresholds $\epsilon_n$ which divide the supercritical and subcritical configurations. The type \uppercase\expandafter{\romannumeral2} critical phenomenon was confirmed for the supercritical configurations and the masses of the BHs follow a power-law behavior~\cite{Husain:2000vm,Pretorius:2000yu,Bizon:2011gg,Jalmuzna:2011qw}.  For the subcritical configurations, a new power-law behavior: $M_{AH}-M_g\propto (\epsilon_c-\epsilon)^\xi$ was first observed recently in \cite{Olivan:2015fmy,Santos-Olivan:2016djn}, where $\xi=0.7$ is universal in the same sense as in the Choptuik's scaling law, 
$M_{AH}$ is the initial black hole mass at the subcritical solution and $M_g$ is the mass gap. Contrary to the  cases discovered by Choptuik and his cooperators which have been understood very well,  this new gapped scaling behavior is still mysterious. For example, we still don't know if there is any asymptotic scaling symmetry near the critical point as the case  found in the  Choptuik's Type \uppercase\expandafter{\romannumeral2} critical phenomena, and whether it can appear in other matter fields and  gravity theories such as Yang-Mills field, Gauss-Bonnet theory and so on.  Note that the critical exponent $\gamma$ in the Choptuik's scaling law is universal for the asymptotically flat and AdS cases.  It would be 
very interesting to see whether the critical exponent $\xi$ in the gapped scaling law is  universal or not in confined spacetimes.

In order to see whether the turbulent behavior is an exclusive domain of asymptotically AdS spacetime or a typical feature of ``confined" Einstein's gravity with reflecting boundary condition, Maliborski \cite{Maliborski:2012gx} investigated the collapse of a massless scalar field enclosed in a cavity. He observed a similar turbulent behavior and multiple critical  phenomena as  in the asymptotically AdS case. The recent work~\cite{Cai:2016yxd} further pointed out that the new gapped critical relationship found in Refs. \cite{Olivan:2015fmy,Santos-Olivan:2016djn} can also appear in asymptotic flat space-time with reflecting wall. One of interesting results reported in Ref. \cite{Cai:2016yxd} is the critical exponent in the mass scaling law with a gap is 0.6, which is different from its value in the asymptotic AdS case reported by Refs. \cite{Olivan:2015fmy,Santos-Olivan:2016djn}. This difference gives a very important signal that this new gapped critical behavior has some essential difference compared with what we have known in  the Choptuik's type \uppercase\expandafter{\romannumeral2} critical behavior, as it has been proven that the critical exponent is independent of the cosmological constant.

To understand why such difference happens and what are the roles of cosmological constant and reflecting wall in this difference, one can investigate such gapped critical behavior with a cosmological constant $\Lambda$ and a reflecting wall located in the radius $R$. Such a setup was first proposed in Refs.  \cite{Buchel:2012uh,Buchel:2013uba,Okawa:2015xma} to investigate the role played by the fully resonant spectrum of AdS in the turbulent instability. It is found that backgrounds with non-resonant frequencies cannot cause collapse at arbitrarily small frequencies \cite{Okawa:2015xma}. In this paper, we study the model as the same as the one in  Refs. \cite{Buchel:2012uh,Buchel:2013uba,Okawa:2015xma}. However, we focus on the critical phenomena near the threshold of black hole formation, instead of turbulent instability of the spacetime. One of our main purposes is to  make a bridge to understand  the difference between the results in the confined asymptotic flat case and  those in the asymptotic AdS case. To be specific, our main motivation is to investigate the influence of cosmological constant and the position of wall on the exponent of mass scaling of the subcritical configurations, which will be shown in this paper that only the value of $\Lambda R^2$ is relevant. Our numerical results show a clear change of critical exponent $\xi$ from 0.37 in  the asymptotic dS case with $\Lambda R^2=1.5$ to 0.7 in the asymptotic AdS case with $\Lambda R^2 \to -\infty$ found in \cite{Olivan:2015fmy,Santos-Olivan:2016djn}. This result is very different from the case in type \uppercase\expandafter{\romannumeral2} mass scaling of supercritical configurations which has already been proven to be independent of the cosmological constant \cite{Hod:1996ar}. In addition, we also find a new time scaling $T_{AH}-T_g\propto(\epsilon_c-\epsilon)^\zeta$ for the forming time of the gapped black hole, where the critical exponent is also dependent of the value of $\Lambda R^2$.

Our paper is organized as follows: In Sec.\ref{sec:level1_2}, we present the equations of motion in double null coordinates and introduce the algorithms briefly. In Sec.\ref{sec:level1_3}, we display the results of our numerical simulations. We conclude in Sec.\ref{sec:level1_4}.

\section{\label{sec:level1_2}Set up}

We consider the gravitational collapse of a real scalar field in a spherical cavity defined by setting a perfectly reflecting mirror at some finite radial radius $r=R$. The dynamics of the system is governed by the Einstein-scalar equations,
\begin{align}
		& G_{\alpha\beta}+\Lambda g_{\alpha\beta}=8\pi G \left[\nabla_\alpha\phi \nabla_\beta\phi-\frac{1}{2} g_{\alpha\beta}(\nabla\phi)^2 \right],\label{eq:Einstein}\\
		& g^{\alpha\beta}\nabla_\alpha \nabla_\beta \phi=0\, \label{eq:scalar} ,
\end{align}
where $G_{\alpha\beta}$ is the Einstein tensor,  $\Lambda$ is the cosmological constant, $\phi$ is the massless scalar field and $G$ is Newton gravitational constant. In what follows we set $G=1$ for convenience.

Following \cite{Garfinkle:1994jb,Cai:2016yxd}, we take the metric ansatz in the double null coordinate,
\begin{align}
	ds^2=-2g\,r'\,dudv+r^2\,d\Omega^2, \label{eq:ansatz}
\end{align}
where $d\Omega^2$ is the two-dimensional sphere metric, $u$ is an outgoing null coordinate and $v$ is an ingoing null coordinate. We take $u$ as the null time and $v$ as the null spatial coordinate. For any quantity $f$, an overdot $\dot{f}$ stands for  $\frac{\partial f}{\partial u}$, while a prime  $f'$ represents $\frac{\partial f}{\partial v}$.

We introduce two auxiliary fields $h$ and $\bar{g}$ such that,
\begin{align}
	\phi &\equiv \bar{h}= \frac{1}{r}\int_{v_0}^v h\, r'd\tilde{v}, \label{eq:hbar}\\
	\bar{g}&=\frac{1}{r}\int_{v_0}^v (1-\Lambda r^2)g\, r'd\tilde{v}, \label{eq:gbar}
\end{align}
where $v_0$ is defined by $r(u,v_0(u))=0$. From the $(v,v)$ and $(u,v)$ components of the Einstein equations \eqref{eq:Einstein}, we get,
\begin{align}
	g&=\exp \left[4\pi\int_{v_0}^{v}\frac{(h-\bar{h})^2}{r}r'd\tilde{v}\right], \label{eq:g}\\
	\dot{r}&=-\frac{1}{2}\bar{g}. \label{eq:rdot}
\end{align}
The evolution of $h$ can be obtained from the Klein-Gordon equation \eqref{eq:scalar},
\begin{align}
	\dot{h}=\frac{1}{2r}(g-\bar{g})(h-\bar{h}). \label{eq:hdot}
\end{align}

The two boundaries of the computational domain are located at $r(u,v)=0$ and $r(u,v)=R$, respectively.  In the double null coordinates, these two boundaries are dynamic and evolve with $u$. So we need to fix the computational domain first by deleting extra grid points in the center and adding more points on the outer  boundary \cite{Cai:2016yxd}, before we impose boundary conditions.

Near $r=0$, in order to overcome the inaccuracy due to the explicit factor of $1/r$ in the expressions for $\bar{h},\bar{g}$ and $g$, we expand $h$ in a Taylor series in $r$ \cite{Garfinkle:1994jb},
\begin{align}
	h=h_0+h_1r+h_2r^2+O(r^3). \label{eq:exp}
\end{align}
Then the expansions for the rest of the variables in $r$ can be obtained by substituting \eqref{eq:exp} into \eqref{eq:hbar}, \eqref{eq:gbar} and \eqref{eq:g}. The coefficients of the expansions can be expressed in terms of  $h_0,h_1,h_2$. All we need to do is to find out these three coefficients. This is done by fitting the first three values of $h$ to a second-order polynomial  \cite{Hod:1996ar}.

We impose the Dirichlet boundary condition at $r=R$,
\begin{align}
	\phi(u,v)|_{r=R}=0.
\end{align}
The value of $\phi$ at those added points can be given by fitting the last few values of $\phi$ and $\phi(R)$ to a fourth-order polynomial. The rest variables can be obtained by definition.

The Misner-Sharp mass contained within the sphere of radius $r$ is defined as
\begin{align}
M(u,r)=\frac{r}{2}\left(1-\frac{\bar{g}}{g}-\frac{\Lambda r^2}{3}\right) \label{eq:MSM}
\end{align}
To find the initial mass of the black hole, we need to find out  where and when the black hole forms. This can be achieved by using apparent horizon(AH). One can easily find that the position of apparent horizon is given by $\bar{g}/g=0$~\footnote{Because of the decomposition of the $(u,v)$ component of metric in Eq.~\ref{eq:ansatz}, $\bar{g}$ and $g$ are both singular at AH. This leads that $\bar{g}/g$ cannot really reach to zero. We set a threshold value and suppose that AH appears when it is less than this threshold}. Let $r_{AH}$ be the initial apparent horizon radius and $r_g$ is its value at the critical solution, from the Eq. \eqref{eq:MSM} we can easily  see that $\delta M_{AH}=M_{AH}-M_g=(1/2-\Lambda r_g^2/2)\delta r_{AH}=(1/2-\Lambda r_g^2/2)(r_{AH}-r_g)$, so the mass and the radius of initial black hole have the same scaling behavior and  share the same critical exponent.

\section{\label{sec:level1_3}Numerical Results}

We use a 4th-order Runge-Kutta scheme to solve the time evolution equations \eqref{eq:rdot}, \eqref{eq:hdot}.
At every time step, those points whose value of $r$ is negative are removed and those points whose value of $r$ is smaller than $R$ are added. Then the functions $\bar{h},g,\bar{g}$ are calculated in sequence using \eqref{eq:hbar}, \eqref{eq:gbar}, \eqref{eq:g}. A detailed description can be found  in \cite{Cai:2016yxd}.

We choose the  Gaussian-type initial data for the scalar field,
\begin{align}
	\phi(r)=\epsilon \exp\left[-\frac{\tan^2 (\frac{\pi r}{R}-\frac{\pi}{2})}{\sigma^2}\right]
\end{align}
where $\epsilon,\sigma$ are two parameters. For our simulations we set $\sigma=1/8$.

Besides these parameters, there are two other parameters, the cosmological constant $\Lambda$ and wall position $R$. Let's consider a scaling transformation:
\begin{align}
	u\rightarrow\frac{u}{k}, v\rightarrow\frac{v}{k}, r\rightarrow\frac{r}{k}, \Lambda\rightarrow\Lambda k^2, R\rightarrow  \frac{R}{k}
\end{align}
This transformation will not change the forms of Eqs. \eqref{eq:hbar}-\eqref{eq:hdot} and will not change the solution of metric and scalar field.
This scaling property means that for the two parameters $\Lambda$ and $R$, only the value of $\Lambda R^2$ is relevant on the system.This leads that we can fix the wall position and only study  the influence of the cosmological constant on the system.  The AdS limit then can be obtained by two equivalent manners, one is fixing $\Lambda$ to be any negative value and then taking $R$ into $\infty$, the other  is fixing $R$ and taking $\Lambda\rightarrow-\infty$.

In addition, we will treat the cosmological constant as a free parameter in the model rather than constrain it to be negative. As the reflecting wall plays the role to confine the energy, even for the asymptotic dS case, the bounce can also appear and weak turbulence can still appear. When the amplitude is large enough, the wave packet collapses to form an AH on its first implosion. When the amplitude is smaller than some threshold $\epsilon_0$, it is too weak to form an AH on its first implosion. The wave package bounces, travels "instantaneously" to the mirror and is reflected back by the mirror. An AH might form on its second implosion. This scenario repeats again and again as we decrease the amplitude continuously. The critical amplitude  which separates those AHs formed on their $(n-1)$th and $n$th implosions is denoted as $\epsilon_{n-1}$.

However, this picture breaks down when the cosmological constant is too positive. In this case, the cosmological constant is so large that the cosmological horizon will be  located inside the cutoff $r=R$, the bounced subcritical wave package can't touch the mirror and can't be reflected back to start its second implosion. This is like  the case  that this mirror does  not exist.  In this case, there is no black hole formation. Therefore we will always consider the case with the reflecting wall is inside the cosmological horizon when the cosmological constant 
is positive. 

To study the influence of the rescaled radius $\Lambda R^2$ on the critical behavior, we change the cosmological constant $\Lambda$, with the cut off $R$ fixed (we choose $R=1$ for convenience). Then the AdS limit corresponds to $\Lambda\rightarrow-\infty$. Note the dS limit doesn't need that $\Lambda R^2\rightarrow\infty$, as we have discussed, because of the existence of the  cosmological horizon, it only needs that $\Lambda R^2>\Lambda_{\text{dS}} R^2>0$, where $\Lambda_{\text{dS}}$ is the solution of $\bar{g}(0,r)/g(0,r)|_{r=R}=0$ for given initial scalar field's configuration. When $\epsilon\rightarrow0$, $\Lambda_{\text{dS}}=3/R^2$. For nonzero $\epsilon$, $\Lambda_{\text{dS}}$ is less than $3/R^2$ and depends on the initial configuration.

\subsection{\label{sec:level2_1}Critical amplitude}
\begin{figure}
	\includegraphics[width=0.4\textwidth]{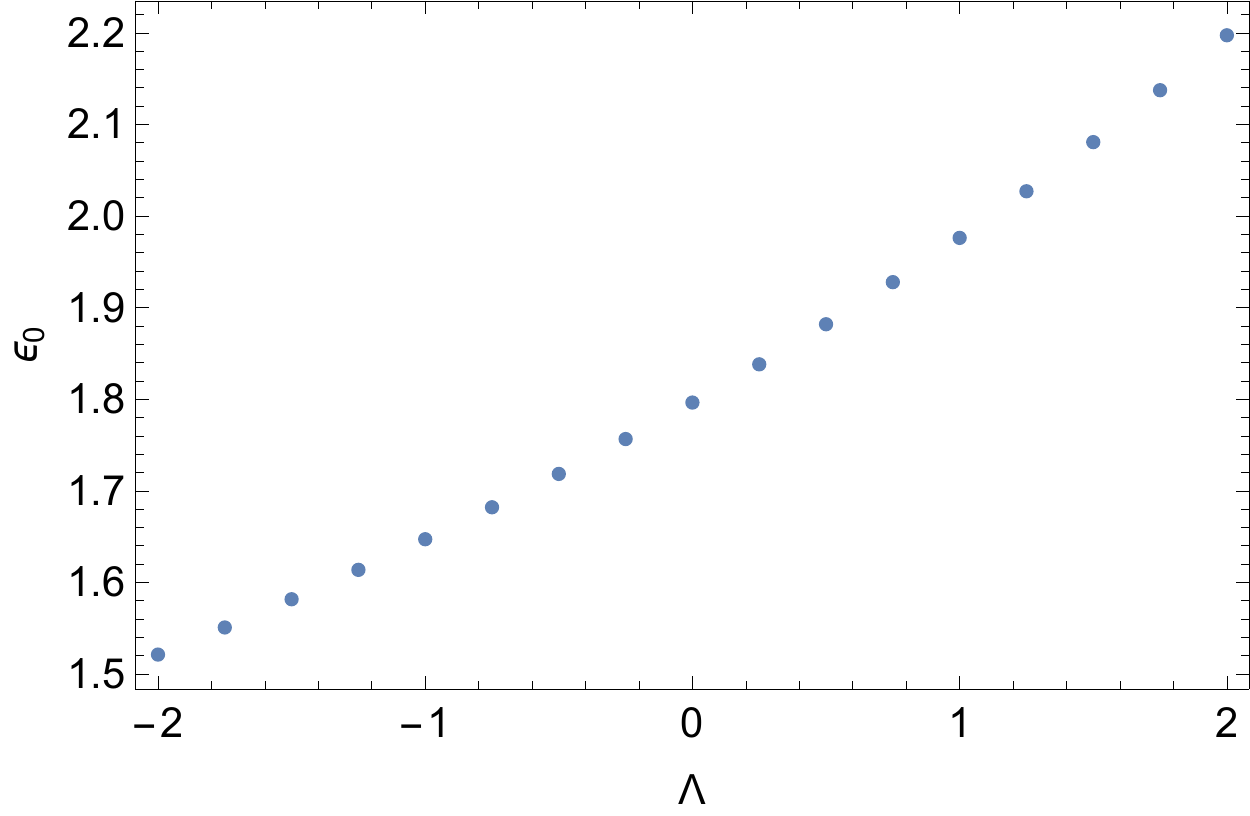}
	\caption{Effect of cosmological constant $\Lambda$ on the first critical amplitude $\epsilon_0$.}
	\label{fig:critical_amp}
\end{figure}

We first consider the effect of cosmological constant $\Lambda$ on the critical amplitudes $\epsilon_n$. For a fixed $\Lambda$, the critical amplitudes $\epsilon_n$ are found using bisection method. The result for $\epsilon_0$ between the first and second branches is shown in Fig.~\ref{fig:critical_amp}.
As we can see, $\epsilon_0$ increases as we increase the cosmological constant $\Lambda$, which indicates that a positive $\Lambda$ suppresses the collapse of  the scalar field, while a negative $\Lambda$ enhances it. This can be understood since a positive cosmological constant provides an additional negative pressure, while a negative cosmological 
constant does the opposite. 

When the magnitude of the cosmological constant is near to zero, the dependence of the critical amplitude on $\Lambda$ is almost linear, which is consistent with the result in \cite{Hod:1996ar}.
	
\subsection{\label{sec:level2_2}Critical black hole}
 Mass gaps between the branches of collapsed scalar fields are also found in our model, which are first noticed by Santos-Oliv\'{a}n and Sopuerta~\cite{Olivan:2015fmy,Santos-Olivan:2016djn} in the asymptotically AdS case. The reason why there are such mass gaps is that the subcritical configurations have to travel to the boundary and come back, suffering from a finite compression, according to~\cite{Olivan:2015fmy,Santos-Olivan:2016djn}. Different  cosmological constant may have different effect on this compression process.

\begin{figure}
	\includegraphics[width=0.45\textwidth]{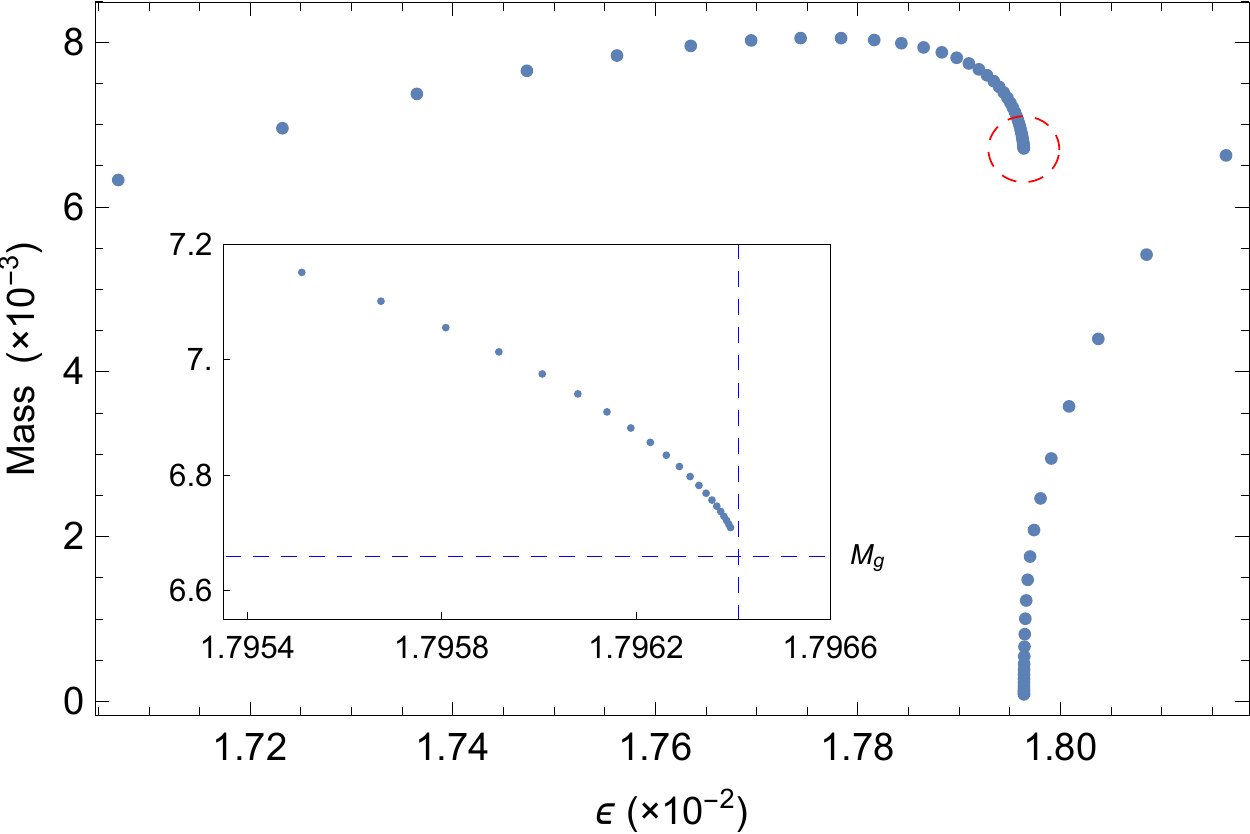}
	\caption{Mass gap between the zero and first bounce branches when $\Lambda=0$. The inset shows the details in the red circle.}
	\label{fig:mgap}
\end{figure}

\begin{figure}
	\includegraphics[width=0.45\textwidth]{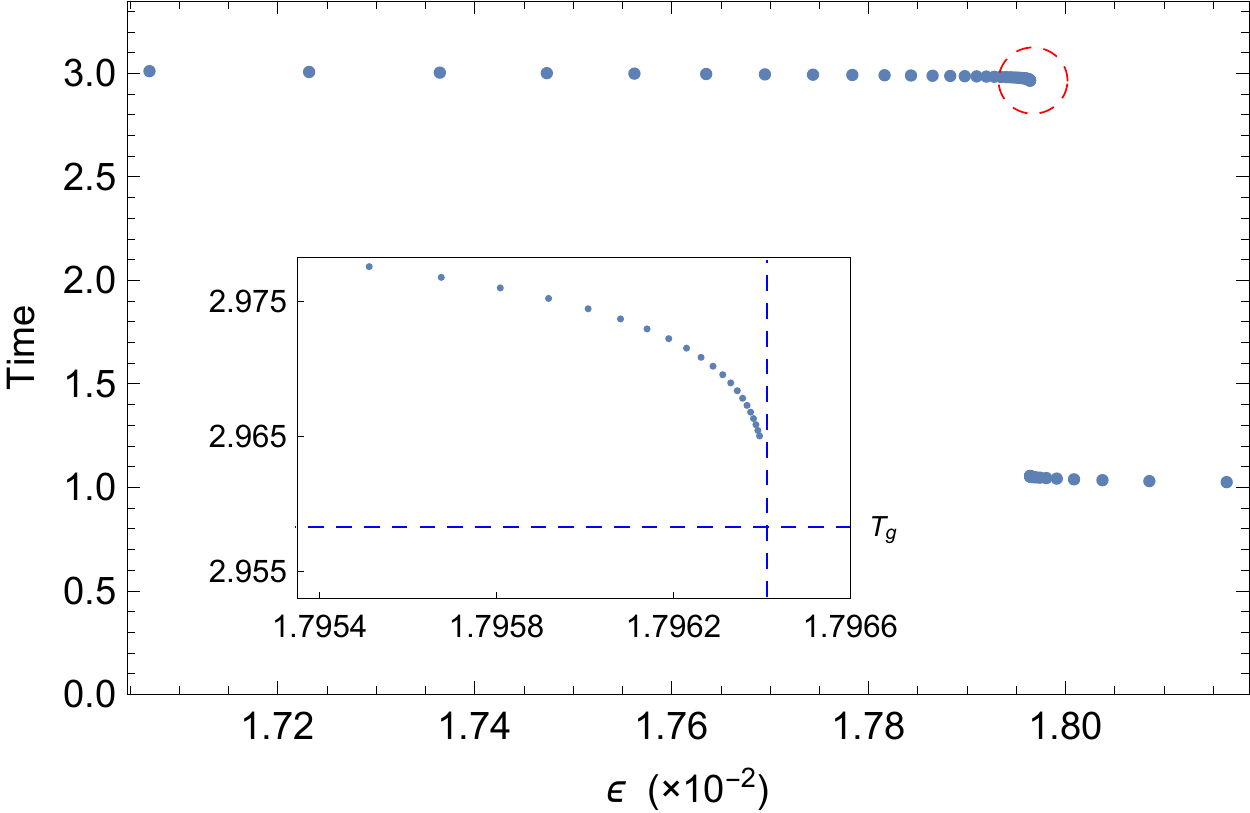}
	\caption{Time gap between the zero and first bounce branches when $\Lambda=0$. The inset shows the details in the red circle. }
	\label{fig:tgap}
\end{figure}

Since the critical amplitudes $\epsilon_n$ for each $\Lambda$ have already been found in the previous section, we can run a simulation for each $\epsilon_n$ to get the critical radius of the black hole and calculate the corresponding Misner-Sharp mass. However, in these cases, the scalar field has already developed very sharp feature before the bounce, which causes great numerical errors during evolution. Instead, we suppose the AH mass follows a power-law  behavior of the type $M_{AH}-M_g^{n+1}\propto (\epsilon_n-\epsilon)^\xi$, following~\cite{Olivan:2015fmy}. We perform a series of subcritical simulations near each critical amplitude, and find the best fitting for $M_g^{n+1}$ and $\xi$ (see Fig.~\ref{fig:mgap}). Interestingly, the collapse time near each critical amplitude is also found to be well fitted by a power-law of the type 
$T_{AH}-T_g^{n+1}\propto(\epsilon_n-\epsilon)^\zeta$. The time gaps $T_g^{n+1}$ and exponent $\zeta$ can be calculated in a similar way  (see Fig.~\ref{fig:tgap}) Since the exponent $\xi$ is universal (in the sense that it is the same for all the mass gaps/branches~\cite{Santos-Olivan:2016djn} and it is also confirmed in our case), here we only focus on the critical behavior near the first mass gap $M_g^1$ for simplicity.

\subsubsection{\label{sec: level3_1}Power-law of black-hole mass}

\begin{figure}
 	\includegraphics[width=0.45\textwidth]{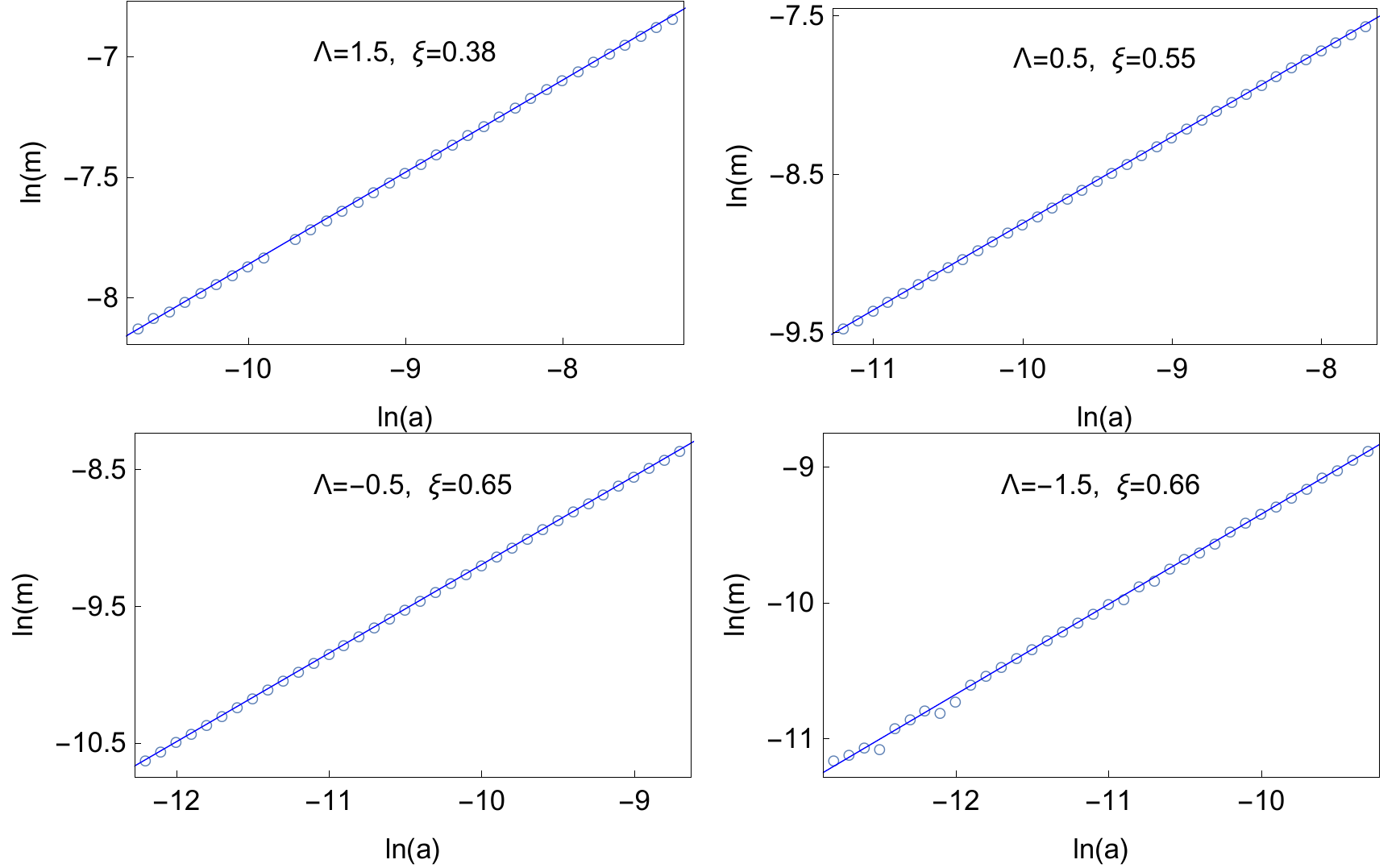}
 	\caption{The black hole mass scaling: $\ln(m)$ vs $\ln(a)$ for different $\Lambda$, where $m=M_{AH}-M_g^1$ and $a=(\epsilon_0-\epsilon)/\epsilon_0$. The points are 
	well fitted by a straight line whose slope $\xi$ increases as we decrease the cosmological constant $\Lambda$. }
 	\label{fig:rscaling}
 \end{figure}

Fig.~\ref{fig:rscaling} shows $\ln (m)$ as a function of $\ln (a)$, where $a=(\epsilon_0-\epsilon)/\epsilon_0$ and $m=M_{AH}-M_g^1$. We vary the cosmological constant $\Lambda$ from $-1.75$ to $1.75$ with the position of the mirror fixed ($R=1$). As we can see, the power-law scaling of masse is clear for each $\Lambda$. We show the results for the fittings in Table \ref{tab:fit_r}.

\begin{table}
	\centering
	\begin{tabular}{r @{.} l *{2}{c}}\hline
		\multicolumn{2}{c}{$\Lambda$} &Mass gap ($M_g^1\times 10^3$) & Exponent($\xi$)\\\hline
		1&75     & 6.8$\pm$0.2        & 0.37$\pm$0.07 \\
		1&50     & 6.95$\pm$0.05   & 0.38$\pm$0.04\\
		1&25     & 7.03$\pm$0.01    & 0.42$\pm$0.02 \\
		1&00     & 7.02$\pm$0.01    & 0.47$\pm$0.02 \\
		0&75     & 6.966$\pm$0.007    & 0.51$\pm$0.02 \\
		0&50     & 6.875$\pm$0.006    & 0.55$\pm$0.02\\
		0&25     & 6.773$\pm$0.003    & 0.58$\pm$0.02 \\
		0&         & 6.665$\pm$0.003    & 0.61$\pm$0.02  \\
		-0&25   & 6.553$\pm$0.001    & 0.63$\pm$0.02  \\
		-0&50   & 6.442$\pm$0.001    & 0.65$\pm$0.01  \\
		-0&75   & 6.330$\pm$0.002    & 0.65$\pm$0.02 \\
		-1&00   & 6.223$\pm$0.001     & 0.68$\pm$0.01  \\
		-1&25   & 6.117$\pm$0.001      & 0.69$\pm$0.01   \\
		-1&50   & 6.011$\pm$0.002     & 0.66$\pm$0.03  \\
		-1&75   & 5.912$\pm$0.001     & 0.68$\pm$0.02   \\
		\hline
	\end{tabular}
	\caption{Fitting data of Fig.~\ref{fig:rscaling} to the power-law: $M_{AH}-M_g^1\propto(\epsilon_0-\epsilon)^\xi$.}
	\label{tab:fit_r}
\end{table}

In~\cite{Hod:1996ar}, Hod and Piran showed that the exponent $\gamma$ of Choptuik's mass scaling: $M_{AH}\propto (\epsilon-\epsilon_0)^\gamma$ for supercritical configurations is immune to the existence of cosmological constant $\Lambda$. It can be understood as follows. The critical solution which determines $\gamma$ shows its structure on smaller and smaller spatial (and temporal) scales as it evolves. And the effect of cosmological constant $\Lambda$ on the critical solution becomes less and less important and ultimately disappears when a naked singularity forms. So the exponent $\gamma$ is expected to be independent of $\Lambda$. However, the situation is different when it comes to the mass scaling on the left side of the mass gaps. The critical solution which determines the exponent $\xi$ collapses into a black hole of a finite size, instead of a naked singularity. So the cosmological constant $\Lambda$ always has a finite contribution. We expect the exponent $\xi$ to be different when $\Lambda$ is different, which is conformed in Table~\ref{tab:fit_r}.

The exponent $\xi$ grows as we decrease the cosmological constant $\Lambda$. We expect it to approach the exponent in the AdS limit which is around $0.70$~\cite{Santos-Olivan:2016djn}, as $\Lambda$ goes to $-\infty$. As we can see in Table~\ref{tab:fit_r}, the precision in the simulation is not so good when the magnitude of $\Lambda$ is large. This is due to the fast adding or deleting grid points in these cases, which causes extra numerical errors.

\begin{figure}
 	\includegraphics[width=0.4\textwidth]{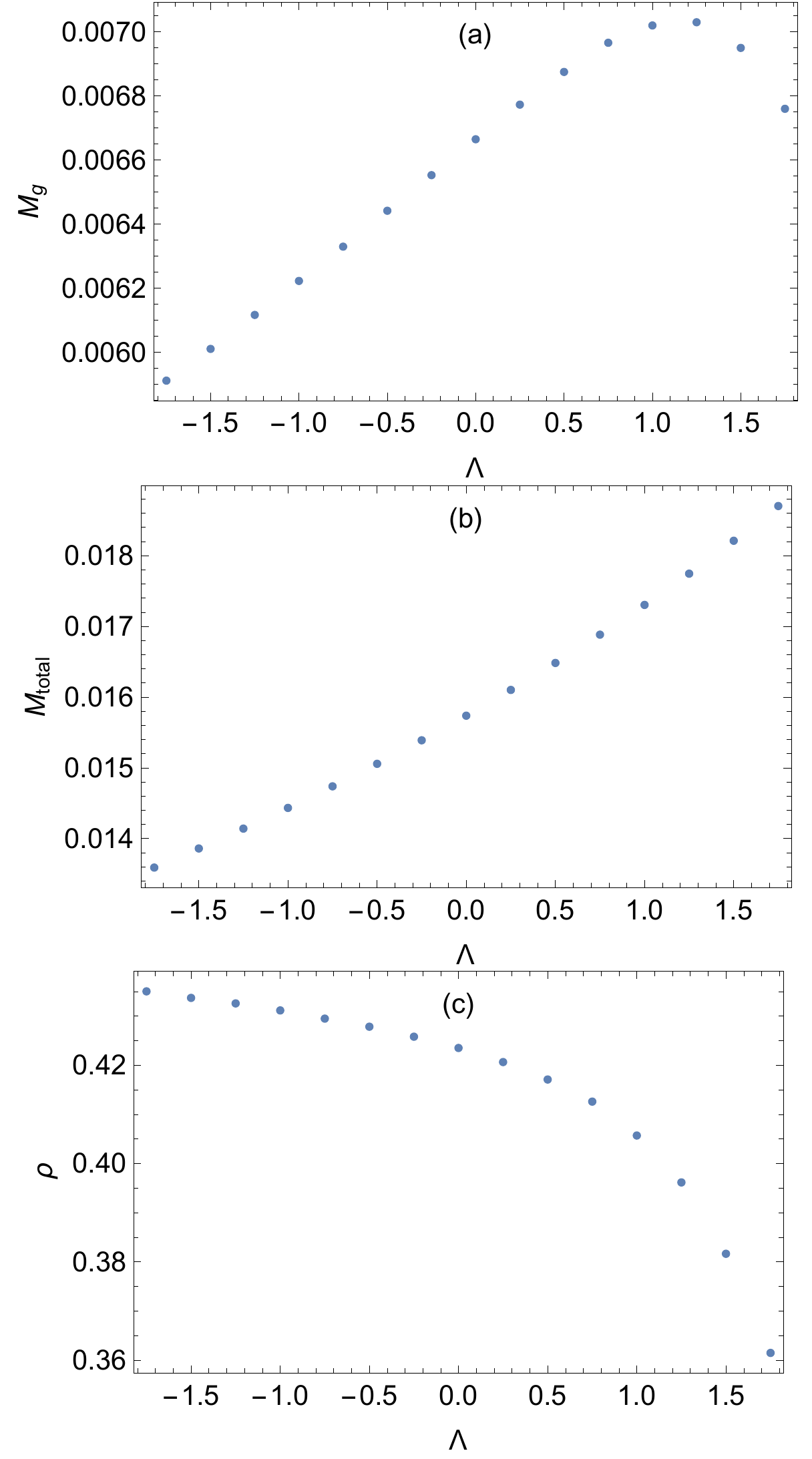}
 	\caption{Masses for different cosmological constant $\Lambda$. Top: Mass gap $M_g^1$, Middle: Total mass of the configuration which collapses into a black hole with masse $M_g^1$, Bottom: The mass ratio $\rho=M_g^1/M_{total}$}
 	\label{fig:cm}
\end{figure}

The mass gap $M_g^1$ corresponds to the mass of critical black holes into which the first subcritical configuration collapses. We plot the mass gaps $M_g^1$ for different $\Lambda$ in Fig.~\ref{fig:cm}(a).It shows a very interesting behavior. We observe a nearly linear increasing behavior of $M_g^1$ as we increase $\Lambda$ in the region $\Lambda \in [-1.75,0.5]$. Out of this region, the growth rate of $M_g^1$ decreases and it starts to shrink around $\Lambda=1.25$.

To understand such a behavior, one should first note that the total mass of the subcritical configuration for each $\Lambda$ is different. The total mass is calculated using (\ref{eq:MSM}). As we can see in Fig.~\ref{fig:cm}(b), it grows as we increase the cosmological constant $\Lambda$, which dominates  the behavior of $M_g^1$ when $\Lambda\in[-1.75,0.5]$. However, in fact there is also a contract effect when we increase the cosmological constant, which may lead mass gap to decrease. Because a negative (positive) $\Lambda$ introduces an enhancement (suppression) to the finite compression of the subcritical configuration during its last travel to the boundary and come back, which makes the scalar field collapse to form an AH a little bit earlier (later). Such effects may lead the position of the AH formed is further from the origin when $\Lambda$ is decreased. In other words, the mass gap $M_g^1$ may be decreased when cosmological constant is increased. To separate the effects coming from the total mass and see  this effect indeed happens clearly, we consider the mass ratio $\rho\equiv M_g/M_{total}$, which describes the ratio of black hole mass to the total mass. The dependence of $\rho$ on $\Lambda$ is displayed Fig.~\ref{fig:cm}(c). The result is consistent with the argument we just give above: $\rho$ is negatively correlated to $\Lambda$. So the behavior of mass gap is just the competition of this two effects and there is indeed an inflexion point shown in  Fig.~\ref{fig:cm}(a).

 Fig. \ref{fig:cm}(c) shows a monotonous dependence of $\Lambda$. Two extremal case that $\Lambda\rightarrow-\infty$ and $\Lambda\rightarrow\Lambda_{\text{dS}}$ are very interesting, though our numerical solver can't directly explore them. For the former case, the system should recover to the asymptotic AdS case, where the value of $M_g/M_{total}$ is a finite nonzero value. For the latter one, when $\Lambda>\Lambda_{\text{dS}}$, cosmological horizon is now located inside the mirror, so the energy will be absorbed into cosmological horizon and the subcritical configuration can not be reflected on the boundary to start its second implosion. When $\Lambda\rightarrow\Lambda_{\text{dS}}$, as the wall is near to the cosmological horizon, one can expect the most part of the energy will be holden by the reflecting wall and the initial black hole mass will  contain a little part of the total energy. Then we have a very interesting conjecture that $\rho$ will approach to some finite value $\rho_0$ as $\Lambda$ goes to $-\infty$ and $\rho\rightarrow0$ at some finite $\Lambda\rightarrow\Lambda_{\text{dS}}$.

\subsubsection{\label{sec: level3_2}Power-law of collapse time}

\begin{figure}
	\includegraphics[width=0.45\textwidth]{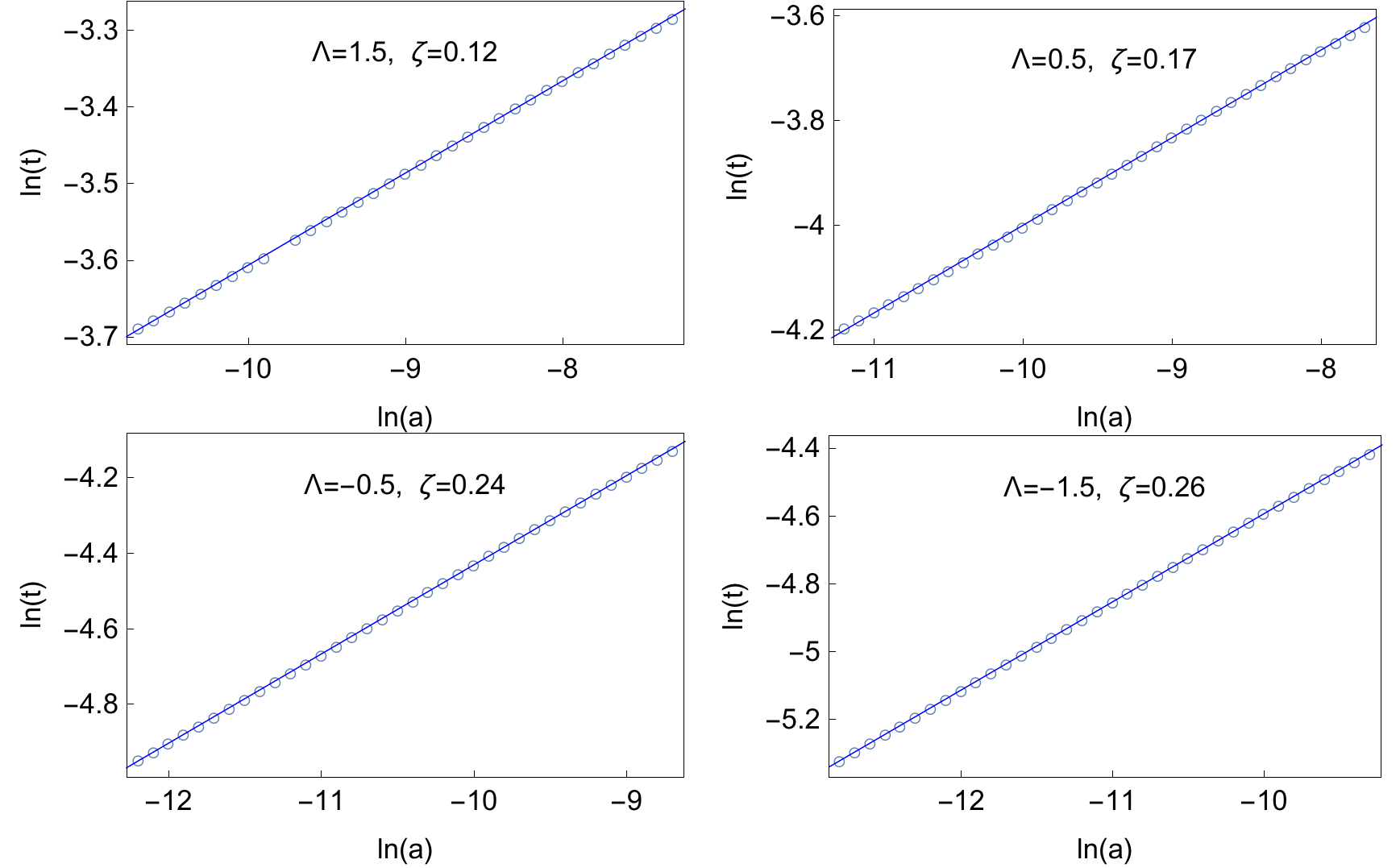}
	\caption{The time scaling: $\ln(t)$ vs $\ln(a)$ for different $\Lambda$, where $t=T_{AH}-T_g^1$ and $a=(\epsilon_0-\epsilon)/\epsilon_0$. The points are well fitted by a straight line whose slope $\zeta$ increases as we decrease the cosmological constant $\Lambda$.}
	\label{fig:tscaling}
\end{figure}

The step-like increase of collapse time is common when one considers the gravitational collapse of some matter fields  in bounded domains~\cite{Bizon:2011gg,Jalmuzna:2011qw,Maliborski:2012gx}. We confirm this structure of collapse time in our simulations. What is more interesting is that when we make a more close look at  the collapse time near the critical amplitude,  we find a power-law behavior: $T_{AH}-T_g^{1}\propto (\epsilon_0-\epsilon)^\zeta$, similar to the power-law relation of the black hole mass. Fig.~\ref{fig:tscaling} shows $\ln(t)$ as a function of $\ln(a)$, where $t=T_{AH}-T_g^1$. The points are well fitted by a straight line whose slope is $\zeta$ for each $\Lambda$. The fitting results are displayed in Table \ref{tab:fit_t}.

The exponent $\zeta$ also increases as we decrease the cosmological constant $\Lambda$. Since the collapse time is less sensitive to the numerical errors than the collapse mass of the black hole, the precision of the parameters displayed in Table \ref{tab:fit_t} is better than the one displayed in Table \ref{tab:fit_r}.

\begin{table}
	\centering
	\begin{tabular}{r @{.} l *{2}{c}}\hline
		\multicolumn{2}{c}{$\Lambda$} & Time gap ($T_g$) & Exponent($\zeta$)\\
		\hline
		1&75     & 3.620$\pm$0.006   &0.10$\pm$0.01\\
		1&50     & 3.476$\pm$0.007   &0.12$\pm$0.02\\
		1&25     & 3.357$\pm$0.001   &0.14$\pm$0.01\\
		1&00     & 3.253$\pm$0.002   &0.14$\pm$0.01\\
		0&75     & 3.166$\pm$0.003 &0.16$\pm$0.02\\
		0&50     & 3.087$\pm$0.001  &0.17$\pm$0.01\\
		0&25     & 3.018$\pm$0.001   &0.19$\pm$0.01\\
		0&         & 2.954$\pm$0.002  &0.20$\pm$0.02\\
		-0&25   & 2.897$\pm$0.001  &0.22$\pm$0.01\\
		-0&50   & 2.844$\pm$0.001  &0.24$\pm$0.01\\
		-0&75   & 2.795$\pm$0.002  &0.24$\pm$0.03\\
		-1&00   & 2.750$\pm$0.001  &0.26$\pm$0.01\\
		-1&25   & 2.707$\pm$0.001   &0.27$\pm$0.01\\
		-1&50   & 2.666$\pm$0.002  &0.26$\pm$0.05\\
		-1&75   & 2.629$\pm$0.001   &0.26$\pm$0.01\\
		\hline
	\end{tabular}
	\caption{Fitting data of Fig.~\ref{fig:tscaling} to the power-law: $T_{AH}-T_g^1\propto(\epsilon_0-\epsilon)^\zeta$.}
	\label{tab:fit_t}
\end{table}

\begin{figure}
	\includegraphics[width=0.4\textwidth]{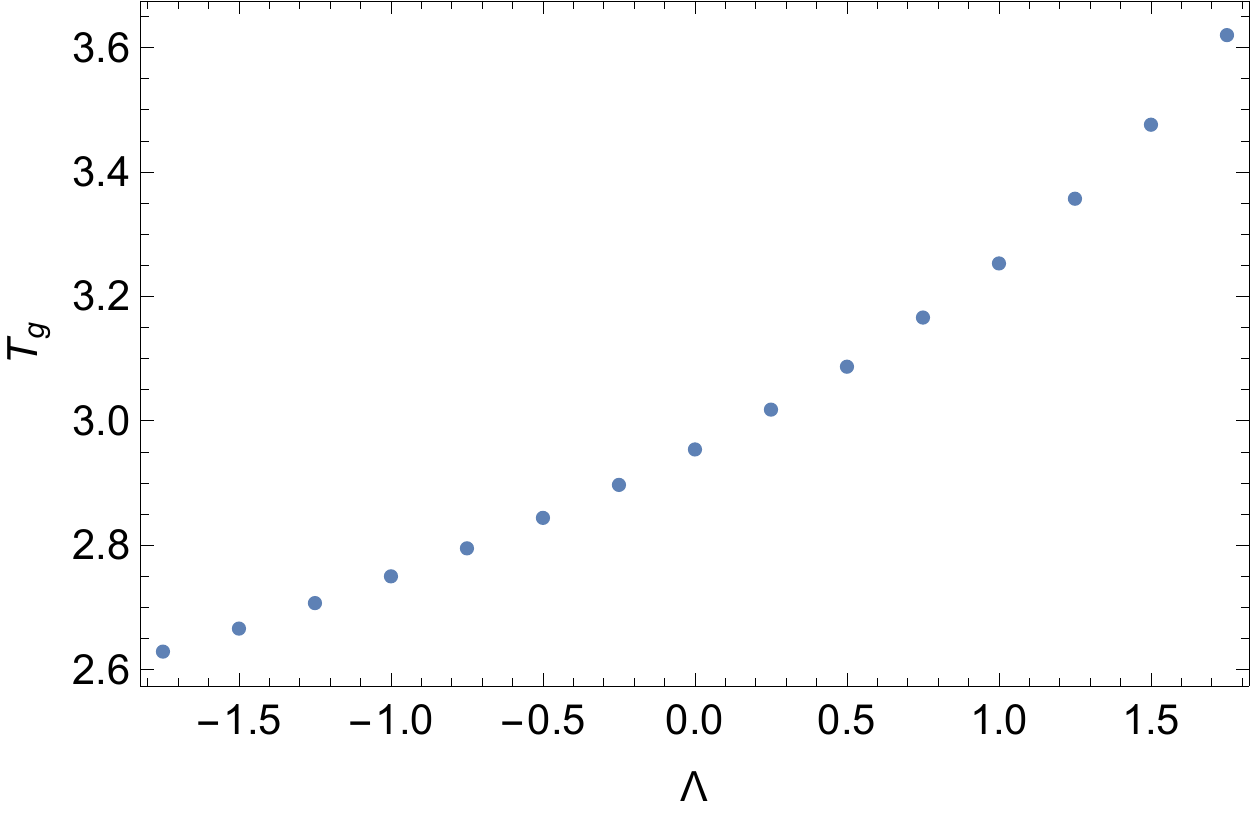}
	\caption{Critical time gap versus different $\Lambda$}
 	\label{fig:ct}
\end{figure}

Fig.~\ref{fig:ct} shows the time gap of critical collapse  $T_g^1$ for different  $\Lambda$. It grows monotonously as we increase $\Lambda$, which is consistent with the argument we gave in the last section: a negative (positive) $\Lambda$ introduces an enhancement (suppression) to the formation of black hole.

As the same as what we have analyzed, we also conjecture that  $T_g^1$ will tend to a finite value when $\Lambda R^2\rightarrow\-\infty$ and tend to infinite when $\Lambda\rightarrow\Lambda_{\text{dS}}$. However, our numerical solver can't give a clear evidence for this conjecture. It is worth further  studying in the future.

\section{\label{sec:level1_4}Conclusion}
Though the confined asymptotic flat space-time and asymptotic AdS space-time share many same properties in black hole forming, such as weak turbulence, multiple type II critical gravitational collapses and so on, an interesting but still poor understood difference among the critical exponents at the new gapped critical point has been pointed out recently. The main aim of this paper is to try make some attempts to understand why such a difference happens and what the roles of cosmological constant and reflecting wall  are in this difference.

 Though the cosmological constant and the position of reflecting wall are both involved, as there is a scaling symmetry, only the value of $\Lambda R^2$ is relevant. Our numerical results showed  a clear evolution of the critical  exponent $\xi$ when we change the value of $\Lambda R^2$. This is a very interesting result and gives out a  clear piece of evidence to show that there are some unknown essential differences compared with  the  Choptuik's type \uppercase\expandafter{\romannumeral2} mass scaling of supercritical configurations which has already been proved to be independent of the cosmological constant in Ref. \cite{Hod:1996ar}. In addition, we also found a new time scaling $T_{AH}-T_g\propto(\epsilon_c-\epsilon)^\zeta$ for the forming time of the gapped black hole, where the critical exponent is also dependent of the value of $\Lambda R^2$.

Note that the Choptuik's type \uppercase\expandafter{\romannumeral2} critical behavior coming from an emergent discrete (or continuous) self-similarity near in the critical region and the critical exponent are related  to the Lyapunov's index \cite{Gundlach:1996eg,Koike:1995jm}. The self-similarity induces a conformal symmetry. The mass and the charge of scalar and the value of cosmological constant are all irrelevant operators in such a conformal transformation. It is interesting to see  whether there is any such a discrete (or continuous) self-similarity in the new gapped critical solution. However, based on the this paper, it becomes  clear that if  such a self-similarity dose exist, the cosmological constant must be a relevant operator. We hope to report further study on this issue in  the future.

\begin{acknowledgements}
This work was finalized during a visit  of R.G. Cai  as a visiting professor to the Yukawa Institute for Theoretical Physics, Kyoto University, the warm hospitality extended to him is greatly appreciated.  This work was supported in part by the  National Natural Science Foundation of China under Grants No.11375247 and No.11435006, and in part by a key project of CAS, Grant No.QYZDJ-SSW-SYS006. 
\end{acknowledgements}

\vskip 5cm

\bibliography{reflect_ref}

\begin{thebibliography}{20}%
\makeatletter
\providecommand \@ifxundefined [1]{%
 \@ifx{#1\undefined}
}%
\providecommand \@ifnum [1]{%
 \ifnum #1\expandafter \@firstoftwo
 \else \expandafter \@secondoftwo
 \fi
}%
\providecommand \@ifx [1]{%
 \ifx #1\expandafter \@firstoftwo
 \else \expandafter \@secondoftwo
 \fi
}%
\providecommand \natexlab [1]{#1}%
\providecommand \enquote  [1]{``#1''}%
\providecommand \bibnamefont  [1]{#1}%
\providecommand \bibfnamefont [1]{#1}%
\providecommand \citenamefont [1]{#1}%
\providecommand \href@noop [0]{\@secondoftwo}%
\providecommand \href [0]{\begingroup \@sanitize@url \@href}%
\providecommand \@href[1]{\@@startlink{#1}\@@href}%
\providecommand \@@href[1]{\endgroup#1\@@endlink}%
\providecommand \@sanitize@url [0]{\catcode `\\12\catcode `\$12\catcode
  `\&12\catcode `\#12\catcode `\^12\catcode `\_12\catcode `\%12\relax}%
\providecommand \@@startlink[1]{}%
\providecommand \@@endlink[0]{}%
\providecommand \url  [0]{\begingroup\@sanitize@url \@url }%
\providecommand \@url [1]{\endgroup\@href {#1}{\urlprefix }}%
\providecommand \urlprefix  [0]{URL }%
\providecommand \Eprint [0]{\href }%
\providecommand \doibase [0]{http://dx.doi.org/}%
\providecommand \selectlanguage [0]{\@gobble}%
\providecommand \bibinfo  [0]{\@secondoftwo}%
\providecommand \bibfield  [0]{\@secondoftwo}%
\providecommand \translation [1]{[#1]}%
\providecommand \BibitemOpen [0]{}%
\providecommand \bibitemStop [0]{}%
\providecommand \bibitemNoStop [0]{.\EOS\space}%
\providecommand \EOS [0]{\spacefactor3000\relax}%
\providecommand \BibitemShut  [1]{\csname bibitem#1\endcsname}%
\let\auto@bib@innerbib\@empty
\bibitem [{\citenamefont {Choptuik}(1993)}]{Choptuik:1992jv}%
  \BibitemOpen
  \bibfield  {author} {\bibinfo {author} {\bibfnamefont {M.~W.}\ \bibnamefont
  {Choptuik}},\ }\href {\doibase 10.1103/PhysRevLett.70.9} {\bibfield
  {journal} {\bibinfo  {journal} {Phys. Rev. Lett.}\ }\textbf {\bibinfo
  {volume} {70}},\ \bibinfo {pages} {9} (\bibinfo {year} {1993})}\BibitemShut
  {NoStop}%
\bibitem [{\citenamefont {Choptuik}\ \emph {et~al.}(1996)\citenamefont
  {Choptuik}, \citenamefont {Chmaj},\ and\ \citenamefont
  {Bizon}}]{Choptuik:1996yg}%
  \BibitemOpen
  \bibfield  {author} {\bibinfo {author} {\bibfnamefont {M.~W.}\ \bibnamefont
  {Choptuik}}, \bibinfo {author} {\bibfnamefont {T.}~\bibnamefont {Chmaj}}, \
  and\ \bibinfo {author} {\bibfnamefont {P.}~\bibnamefont {Bizon}},\ }\href
  {\doibase 10.1103/PhysRevLett.77.424} {\bibfield  {journal} {\bibinfo
  {journal} {Phys. Rev. Lett.}\ }\textbf {\bibinfo {volume} {77}},\ \bibinfo
  {pages} {424} (\bibinfo {year} {1996})},\ \Eprint
  {http://arxiv.org/abs/gr-qc/9603051} {arXiv:gr-qc/9603051 [gr-qc]}
  \BibitemShut {NoStop}%
\bibitem [{\citenamefont {Gundlach}(2003)}]{Gundlach:2002sx}%
  \BibitemOpen
  \bibfield  {author} {\bibinfo {author} {\bibfnamefont {C.}~\bibnamefont
  {Gundlach}},\ }\href {\doibase 10.1016/S0370-1573(02)00560-4} {\bibfield
  {journal} {\bibinfo  {journal} {Phys. Rept.}\ }\textbf {\bibinfo {volume}
  {376}},\ \bibinfo {pages} {339} (\bibinfo {year} {2003})},\ \Eprint
  {http://arxiv.org/abs/gr-qc/0210101} {arXiv:gr-qc/0210101 [gr-qc]}
  \BibitemShut {NoStop}%
\bibitem [{\citenamefont {Gundlach}\ and\ \citenamefont
  {Martin-Garcia}(2007)}]{Gundlach:2007gc}%
  \BibitemOpen
  \bibfield  {author} {\bibinfo {author} {\bibfnamefont {C.}~\bibnamefont
  {Gundlach}}\ and\ \bibinfo {author} {\bibfnamefont {J.~M.}\ \bibnamefont
  {Martin-Garcia}},\ }\href {\doibase 10.12942/lrr-2007-5} {\bibfield
  {journal} {\bibinfo  {journal} {Living Rev. Rel.}\ }\textbf {\bibinfo
  {volume} {10}},\ \bibinfo {pages} {5} (\bibinfo {year} {2007})},\ \Eprint
  {http://arxiv.org/abs/0711.4620} {arXiv:0711.4620 [gr-qc]} \BibitemShut
  {NoStop}%
\bibitem [{\citenamefont {Husain}\ and\ \citenamefont
  {Olivier}(2001)}]{Husain:2000vm}%
  \BibitemOpen
  \bibfield  {author} {\bibinfo {author} {\bibfnamefont {V.}~\bibnamefont
  {Husain}}\ and\ \bibinfo {author} {\bibfnamefont {M.}~\bibnamefont
  {Olivier}},\ }\href {\doibase 10.1088/0264-9381/18/2/101} {\bibfield
  {journal} {\bibinfo  {journal} {Class. Quant. Grav.}\ }\textbf {\bibinfo
  {volume} {18}},\ \bibinfo {pages} {L1} (\bibinfo {year} {2001})},\ \Eprint
  {http://arxiv.org/abs/gr-qc/0008060} {arXiv:gr-qc/0008060 [gr-qc]}
  \BibitemShut {NoStop}%
\bibitem [{\citenamefont {Pretorius}\ and\ \citenamefont
  {Choptuik}(2000)}]{Pretorius:2000yu}%
  \BibitemOpen
  \bibfield  {author} {\bibinfo {author} {\bibfnamefont {F.}~\bibnamefont
  {Pretorius}}\ and\ \bibinfo {author} {\bibfnamefont {M.~W.}\ \bibnamefont
  {Choptuik}},\ }\href {\doibase 10.1103/PhysRevD.62.124012} {\bibfield
  {journal} {\bibinfo  {journal} {Phys. Rev.}\ }\textbf {\bibinfo {volume}
  {D62}},\ \bibinfo {pages} {124012} (\bibinfo {year} {2000})},\ \Eprint
  {http://arxiv.org/abs/gr-qc/0007008} {arXiv:gr-qc/0007008 [gr-qc]}
  \BibitemShut {NoStop}%
\bibitem [{\citenamefont {Bizon}\ and\ \citenamefont
  {Rostworowski}(2011)}]{Bizon:2011gg}%
  \BibitemOpen
  \bibfield  {author} {\bibinfo {author} {\bibfnamefont {P.}~\bibnamefont
  {Bizon}}\ and\ \bibinfo {author} {\bibfnamefont {A.}~\bibnamefont
  {Rostworowski}},\ }\href {\doibase 10.1103/PhysRevLett.107.031102} {\bibfield
   {journal} {\bibinfo  {journal} {Phys. Rev. Lett.}\ }\textbf {\bibinfo
  {volume} {107}},\ \bibinfo {pages} {031102} (\bibinfo {year} {2011})},\
  \Eprint {http://arxiv.org/abs/1104.3702} {arXiv:1104.3702 [gr-qc]}
  \BibitemShut {NoStop}%
\bibitem [{\citenamefont {Jalmuzna}\ \emph {et~al.}(2011)\citenamefont
  {Jalmuzna}, \citenamefont {Rostworowski},\ and\ \citenamefont
  {Bizon}}]{Jalmuzna:2011qw}%
  \BibitemOpen
  \bibfield  {author} {\bibinfo {author} {\bibfnamefont {J.}~\bibnamefont
  {Jalmuzna}}, \bibinfo {author} {\bibfnamefont {A.}~\bibnamefont
  {Rostworowski}}, \ and\ \bibinfo {author} {\bibfnamefont {P.}~\bibnamefont
  {Bizon}},\ }\href {\doibase 10.1103/PhysRevD.84.085021} {\bibfield  {journal}
  {\bibinfo  {journal} {Phys. Rev.}\ }\textbf {\bibinfo {volume} {D84}},\
  \bibinfo {pages} {085021} (\bibinfo {year} {2011})},\ \Eprint
  {http://arxiv.org/abs/1108.4539} {arXiv:1108.4539 [gr-qc]} \BibitemShut
  {NoStop}%
\bibitem [{\citenamefont {Santos-Oliván}\ and\ \citenamefont
  {Sopuerta}(2016{\natexlab{a}})}]{Olivan:2015fmy}%
  \BibitemOpen
  \bibfield  {author} {\bibinfo {author} {\bibfnamefont {D.}~\bibnamefont
  {Santos-Oliván}}\ and\ \bibinfo {author} {\bibfnamefont {C.~F.}\ \bibnamefont
  {Sopuerta}},\ }\href {\doibase 10.1103/PhysRevLett.116.041101} {\bibfield
  {journal} {\bibinfo  {journal} {Phys. Rev. Lett.}\ }\textbf {\bibinfo
  {volume} {116}},\ \bibinfo {pages} {041101} (\bibinfo {year}
  {2016}{\natexlab{a}})},\ \Eprint {http://arxiv.org/abs/1511.04344}
  {arXiv:1511.04344 [gr-qc]} \BibitemShut {NoStop}%
\bibitem [{\citenamefont {Santos-Oliván}\ and\ \citenamefont
  {Sopuerta}(2016{\natexlab{b}})}]{Santos-Olivan:2016djn}%
  \BibitemOpen
  \bibfield  {author} {\bibinfo {author} {\bibfnamefont {D.}~\bibnamefont
  {Santos-Oliván}}\ and\ \bibinfo {author} {\bibfnamefont {C.~F.}\ \bibnamefont
  {Sopuerta}},\ }\href {\doibase 10.1103/PhysRevD.93.104002} {\bibfield
  {journal} {\bibinfo  {journal} {Phys. Rev.}\ }\textbf {\bibinfo {volume}
  {D93}},\ \bibinfo {pages} {104002} (\bibinfo {year} {2016}{\natexlab{b}})},\
  \Eprint {http://arxiv.org/abs/1603.03613} {arXiv:1603.03613 [gr-qc]}
  \BibitemShut {NoStop}%
\bibitem [{\citenamefont {Maliborski}(2012)}]{Maliborski:2012gx}%
  \BibitemOpen
  \bibfield  {author} {\bibinfo {author} {\bibfnamefont {M.}~\bibnamefont
  {Maliborski}},\ }\href {\doibase 10.1103/PhysRevLett.109.221101} {\bibfield
  {journal} {\bibinfo  {journal} {Phys. Rev. Lett.}\ }\textbf {\bibinfo
  {volume} {109}},\ \bibinfo {pages} {221101} (\bibinfo {year} {2012})},\
  \Eprint {http://arxiv.org/abs/1208.2934} {arXiv:1208.2934 [gr-qc]}
  \BibitemShut {NoStop}%
\bibitem [{\citenamefont {Cai}\ and\ \citenamefont {Yang}(2016)}]{Cai:2016yxd}%
  \BibitemOpen
  \bibfield  {author} {\bibinfo {author} {\bibfnamefont {R.-G.}\ \bibnamefont
  {Cai}}\ and\ \bibinfo {author} {\bibfnamefont {R.-Q.}\ \bibnamefont {Yang}},\
  }\href@noop {} {\  (\bibinfo {year} {2016})},\ \Eprint
  {http://arxiv.org/abs/1602.00112} {arXiv:1602.00112 [gr-qc]} \BibitemShut
  {NoStop}%
\bibitem [{\citenamefont {Buchel}\ \emph {et~al.}(2012)\citenamefont {Buchel},
  \citenamefont {Lehner},\ and\ \citenamefont {Liebling}}]{Buchel:2012uh}%
  \BibitemOpen
  \bibfield  {author} {\bibinfo {author} {\bibfnamefont {A.}~\bibnamefont
  {Buchel}}, \bibinfo {author} {\bibfnamefont {L.}~\bibnamefont {Lehner}}, \
  and\ \bibinfo {author} {\bibfnamefont {S.~L.}\ \bibnamefont {Liebling}},\
  }\href {\doibase 10.1103/PhysRevD.86.123011} {\bibfield  {journal} {\bibinfo
  {journal} {Phys. Rev.}\ }\textbf {\bibinfo {volume} {D86}},\ \bibinfo {pages}
  {123011} (\bibinfo {year} {2012})},\ \Eprint {http://arxiv.org/abs/1210.0890}
  {arXiv:1210.0890 [gr-qc]} \BibitemShut {NoStop}%
\bibitem [{\citenamefont {Buchel}\ \emph {et~al.}(2013)\citenamefont {Buchel},
  \citenamefont {Liebling},\ and\ \citenamefont {Lehner}}]{Buchel:2013uba}%
  \BibitemOpen
  \bibfield  {author} {\bibinfo {author} {\bibfnamefont {A.}~\bibnamefont
  {Buchel}}, \bibinfo {author} {\bibfnamefont {S.~L.}\ \bibnamefont
  {Liebling}}, \ and\ \bibinfo {author} {\bibfnamefont {L.}~\bibnamefont
  {Lehner}},\ }\href {\doibase 10.1103/PhysRevD.87.123006} {\bibfield
  {journal} {\bibinfo  {journal} {Phys. Rev.}\ }\textbf {\bibinfo {volume}
  {D87}},\ \bibinfo {pages} {123006} (\bibinfo {year} {2013})},\ \Eprint
  {http://arxiv.org/abs/1304.4166} {arXiv:1304.4166 [gr-qc]} \BibitemShut
  {NoStop}%
\bibitem [{\citenamefont {Okawa}\ \emph {et~al.}(2015)\citenamefont {Okawa},
  \citenamefont {Lopes},\ and\ \citenamefont {Cardoso}}]{Okawa:2015xma}%
  \BibitemOpen
  \bibfield  {author} {\bibinfo {author} {\bibfnamefont {H.}~\bibnamefont
  {Okawa}}, \bibinfo {author} {\bibfnamefont {J.~C.}\ \bibnamefont {Lopes}}, \
  and\ \bibinfo {author} {\bibfnamefont {V.}~\bibnamefont {Cardoso}},\
  }\href@noop {} {\  (\bibinfo {year} {2015})},\ \Eprint
  {http://arxiv.org/abs/1504.05203} {arXiv:1504.05203 [gr-qc]} \BibitemShut
  {NoStop}%
\bibitem [{\citenamefont {Hod}\ and\ \citenamefont {Piran}(1997)}]{Hod:1996ar}%
  \BibitemOpen
  \bibfield  {author} {\bibinfo {author} {\bibfnamefont {S.}~\bibnamefont
  {Hod}}\ and\ \bibinfo {author} {\bibfnamefont {T.}~\bibnamefont {Piran}},\
  }\href {\doibase 10.1103/PhysRevD.55.3485} {\bibfield  {journal} {\bibinfo
  {journal} {Phys. Rev.}\ }\textbf {\bibinfo {volume} {D55}},\ \bibinfo {pages}
  {3485} (\bibinfo {year} {1997})},\ \Eprint
  {http://arxiv.org/abs/gr-qc/9606093} {arXiv:gr-qc/9606093 [gr-qc]}
  \BibitemShut {NoStop}%
\bibitem [{\citenamefont {Garfinkle}(1995)}]{Garfinkle:1994jb}%
  \BibitemOpen
  \bibfield  {author} {\bibinfo {author} {\bibfnamefont {D.}~\bibnamefont
  {Garfinkle}},\ }\href {\doibase 10.1103/PhysRevD.51.5558} {\bibfield
  {journal} {\bibinfo  {journal} {Phys. Rev.}\ }\textbf {\bibinfo {volume}
  {D51}},\ \bibinfo {pages} {5558} (\bibinfo {year} {1995})},\ \Eprint
  {http://arxiv.org/abs/gr-qc/9412008} {arXiv:gr-qc/9412008 [gr-qc]}
  \BibitemShut {NoStop}%
\bibitem [{Note1()}]{Note1}%
  \BibitemOpen
  \bibinfo {note} {Because of the decomposition of the $(u,v)$ component of
  metric in Eq.~\ref {eq:ansatz}, $\protect \mathaccentV {bar}016{g}$ and $g$
  are both singular at AH. This leads that $\protect \mathaccentV
  {bar}016{g}/g$ cannot really reach to zero. We set a threshold value and
  suppose that AH appears when it is less than this threshold}\BibitemShut
  {NoStop}%
\bibitem [{\citenamefont {Gundlach}(1997)}]{Gundlach:1996eg}%
  \BibitemOpen
  \bibfield  {author} {\bibinfo {author} {\bibfnamefont {C.}~\bibnamefont
  {Gundlach}},\ }\href {\doibase 10.1103/PhysRevD.55.695} {\bibfield  {journal}
  {\bibinfo  {journal} {Phys. Rev.}\ }\textbf {\bibinfo {volume} {D55}},\
  \bibinfo {pages} {695} (\bibinfo {year} {1997})},\ \Eprint
  {http://arxiv.org/abs/gr-qc/9604019} {arXiv:gr-qc/9604019 [gr-qc]}
  \BibitemShut {NoStop}%
\bibitem [{\citenamefont {Koike}\ \emph {et~al.}(1995)\citenamefont {Koike},
  \citenamefont {Hara},\ and\ \citenamefont {Adachi}}]{Koike:1995jm}%
  \BibitemOpen
  \bibfield  {author} {\bibinfo {author} {\bibfnamefont {T.}~\bibnamefont
  {Koike}}, \bibinfo {author} {\bibfnamefont {T.}~\bibnamefont {Hara}}, \ and\
  \bibinfo {author} {\bibfnamefont {S.}~\bibnamefont {Adachi}},\ }\href
  {\doibase 10.1103/PhysRevLett.74.5170} {\bibfield  {journal} {\bibinfo
  {journal} {Phys. Rev. Lett.}\ }\textbf {\bibinfo {volume} {74}},\ \bibinfo
  {pages} {5170} (\bibinfo {year} {1995})},\ \Eprint
  {http://arxiv.org/abs/gr-qc/9503007} {arXiv:gr-qc/9503007 [gr-qc]}
  \BibitemShut {NoStop}%
\end{thebibliography}%

\end{document}